# SABIA: An AI-Powered Tool for Detecting Opioid-Related Behaviors on Social Media

Muhammad Ahmad[1,*], Fida Ullah[1], Ildar Batyrshin[1], Grigori Sidorov[1]

[1] Centro de Investigación en Computación, Instituto Politécnico Nacional (CIC-PN), Mexico City 07738, Mexico.
Correspondence: mahmad2024@cic.ipn.mx

**Background:** Social media platforms have become valuable tools for understanding public health challenges by providing insights into patient behaviors, medication use, and mental health issues. However, analyzing such data remains challenging due to the prevalence of informal language, slang, and coded communication, which can obscure the detection of opioid misuse.

**Method:** This study addresses the issue of opioid-related user behavior on social media, including informal expressions, slang terms, and misspelled or coded language. To this end, we analyzed the existing Bidirectional Encoder Representations from Transformers (BERT) technique and developed a BERT-BiLSTM-3CNN hybrid deep learning model, named SABIA, to create a single-task classifier that effectively captures the features of the target dataset. We explore the strengths of the SABIA that showed a benchmark in capturing the semantics and contextual information within the content. The proposed model consists of several stages: 1) data preprocessing, 2) representation of data using SABIA model, 3) fine-tuning phase, and 4) classification of user behavior into five different labels. A new dataset for detecting opioid drug user behavior across multiple classes—such as Dealers, Active Opioid Users, Recovered Users, Prescription Users, and Non-Users—was constructed using Reddit posts, along with detailed annotation guidelines. Several experiments are conducted by using supervised learning.

**Results:** The findings show that SABIA achieved benchmark performance, outperforming the baseline (Logistic Regression, LR = 0.86) and improving accuracy by 9.30%. Finally, we compared SABIA with seven previous studies on opioid crisis detection, where it consistently outperformed existing approaches, confirming its effectiveness and robustness.

**Conclusions:** This study highlights the effectiveness of hybrid deep learning models in detecting complex opioid-related behaviors on social media. SABIA provides a promising foundation for real-time monitoring of drug-related discourse and supports efforts in public health surveillance and intervention.

# 1     Introduction

The opioid crisis has become one of the most devastating public health emergencies worldwide, with the United States alone experiencing hundreds of thousands of over-dose-related deaths over the past two decades [1-5]. The epidemic started with the late 1990s over prescription of opioid pain relievers [6-8] and has now shifted to the illicit use of highly potent synthetic opioids like fentanyl [9-11] and carfentanil [12] [13]. With the heightened rate of awareness and control over prescription opioids, many people have found themselves flocking to illegal sources in order to sustain their addiction [14] [15]. One of the most concerning tendencies that has appeared over the past few years is the use of social media to promote the illicit trade and distribution of opioids [16] [17]. Such platforms have unintentionally turned into a market in which dealers place adverts and distribute opioids directly to their customers, using slang words, hashtags, personal messaging, and code words to evade checks.

The proliferation of black-market opioid sales on social media platforms has led to serious public health and societal consequences. The most popular platforms, such as X (former Twitter), Instagram, Reddit and Telegram, have the largest number of young adult and teenage users, which makes them extremely susceptible to exposure and exploitation. Drug dealers often use encrypted language and internet memes as covert marketing strategies on social media [18] [19]. Furthermore, sales and purchases usually occur through private conversations and are settled using digital wallets or crypto-currencies, ensuring privacy [20] [21]. These substances are very accessible and easy to buy online, which not only makes them enticingly tempting, but especially appealing to anyone feeling emotionally or even physically distressed.

In the United States, numerous licensed healthcare professionals have breached their oath, engaging in the illegal sale of opioids—often driven by money, sex, or personal gain. In February 2021, Dr. Morris Brown of Dayton, Ohio, was sentenced to two years in federal prison after he pleaded guilty to the unlawful distribution of opioids via false prescriptions [22]. In November 2023, Dr. Oscar Lightner and his office manager in Texas were found guilty of operating a so-called pill mill, in which prescriptions that supplied more than half a million opioid pills were issued in exchange for cash payments of between 250 and 500 dollars per prescription, totaling 1.2 billion dollars in illegal income [23]. In yet another famous incident in May 2021, which occurred in New York, Dr. Santiamo was sentenced to 57 months in prison after prescribing oxycodone without any medical need and exchanging sexual favors with young patients [24]. In June 2016, a psychiatrist was given an 11- to 22-year sentence after being found guilty of trading pills (opioids) in exchange for sex [25]. Historically, trusted medical workers have sometimes acted like drug dealers, taking advantage of desperate patients to obtain money or sex, further escalating the opioid crisis.

Understanding the reasons behind the use of opioids is an important step toward overcoming this crisis. Many patients start taking opioids with good intentions, seeking treatment for chronic pain due to surgery or injuries [26-28]. The euphoric and stress-relieving effects of opioids, however, may lead to dependence and addiction when consumed over the long term [29]. For many individuals, opioids serve as a means of escape from emotional trauma, untreated mental health conditions such as anxiety, PTSD, or depression, and socioeconomic challenges including homelessness, unemployment, and poverty. These issues were further exacerbated by the COVID-19 pandemic, which isolated vulnerable people and disrupted access to treatment and counseling.

Now a days Natural Language Processing (NLP) has become a revolutionary tool for the health care surveillance, analysis, and prediction of opioid abuse and illicit trafficking. NLP empowers researchers and healthcare professional with the capability to derive imperative information from hundreds of thousands of unstructured data points, making it attainable to track real-time behaviors related to opioids. NLP helps to track the online opioid ecosystem, unveiling meaningful patterns and sentiment [43] [50], as well as identifying keywords, slang terms related to opioids, and enabling highly accurate classification of user behavior. In addition to substance abuse, NLP has been widely applied in many other domains, such as hate speech detection [30], named entity recognition (NER) [31] [42], hope speech [32-34], and mental health monitoring through social media discourse. NLP plays an important role in the healthcare industry, aiding in the text mining of clinical data, medical diagnoses, and analysis of patient reviews. Such different applications demonstrate the usefulness of NLP not only in public health surveillance but also in supporting law enforcement and guiding evidence-based policymaking.

In this study, we sourced a wide range of social media posts from Reddit, a popular and widely used platform for open discussions related to opioid drugs and manually annotated the posts according to predefined multi-class categories. In our dataset, we do not only include the posts that explicitly use such words as opioid, opiate, fentanyl, or heroin; we also consider posts that contain the most common slang tokens such as black, chocolate, oxy, fent, perk, and roxie etc. to cover a larger set of user posts and we also added some commonly misspelled keywords of different drug names.

Following the construction of the annotated dataset, we employed various natural language processing (NLP) methodologies—including machine learning, deep learning, and advanced pretrained language models—to develop highly accurate classifiers. These classifiers are designed to assist policymakers, healthcare professionals, and law enforcement agencies in identifying and monitoring specific opioid-related user behaviors, such as distinguishing between Dealers, Active Opioid Users, Recovered Users, Prescription Users, and Non-Users based on social media activity. Among the approaches, we developed a hybrid model, BERT-BiLSTM-3CNN, which we named SABIA. Sabia is a Spanish word meaning 'wise' or 'intelligent'.

This study makes the following contribution:

- ➢ We manually built a novel multi-class dataset that includes various user behaviors related to opioids use such as, Dealers, Active Opioid Users, Recovered Users, Prescription Users, and Non-Users—by incorporating both formal and slang terminology, including commonly misspelled and code-word variations of drug names.

- ➢ We developed annotation guidelines and described our annotation procedure in detail. Additionally, we discussed key challenges encountered during the construction of the dataset, such as the use of slang, ambiguity, and fragmented sentences.

   ➢  We analyzed the existing Bidirectional Encoder Representations from Transformers (BERT) technique and developed a BERT-BiLSTM-3CNN hybrid deep learning model, named SABIA, to create a single-task classifier that effectively captures the features of the target dataset and handles noisy, informal expressions, slang, and misspellings in social media text related to opioid behaviors.

   ➢  The findings show that SABIA achieved benchmark performance, outperforming the baseline (Logistic Regression, LR = 0.86) and improving accuracy by 9.30%. Finally, we compared SABIA with seven previous studies on opioid crisis detection, where it consistently outperformed existing approaches, confirming its effectiveness and robustness (see Table 1).

The remaining sections of the study are as follows. The Literature Review examines previous research on opioid crises detection on social media. The Methodology and Design section outlines the approach and system design. Findings are detailed in the Results and Analysis section. The Limitations section discusses the constraints of our proposed solution. Finally, the study concludes with Conclusion and Future Work, summarizing the key outcomes and suggesting directions for further research.

## 2      Literature Review

Researchers have increasingly turned to social media platforms, particularly Reddit, to explore opioid use disorder (OUD) and identify at-risk individuals. Several studies have utilized machine learning and natural language processing techniques to classify user behavior and extract meaningful patterns. For instance, Wang et al. [35] collected Reddit posts from over 1,000 users across opioid-related subreddits, employing an attention-based BiLSTM model for binary classification of opioid users versus non-users. While this model not only outperformed traditional methods in terms of F1-score and offered interpretability by highlighting indicative terms through the attention mechanism, our work differs by addressing a multi-class classification problem rather than a binary setting. Building on similar methods, Yao et al. [36] expanded the scope to include multiple subreddit types (suicide, depression, opioid-related, and control), training neural network classifiers—particularly convolutional ones—to identify suicidal content in the context of opioid use. Their findings underscore the potential of machine learning to detect individuals who may be struggling but not actively seeking help.

Beyond classification, other studies have focused on understanding user behavior, information needs, and the social support aspects of online communities. Laud et al. [38], for example, analyzed OUD-related questions on Reddit using transformer-based models and hierarchical clustering. They categorized posts into ten major themes—such as withdrawal, treatment, and drug testing—offering valuable insights into what information people with OUD seek online. Similarly, Mittal et al. [39] explored how myths about OUD are circulated and challenged within Reddit posts and LLM-generated responses. They found that human responses often perpetuated myths using authoritative language, whereas myth-correcting messages tended to adopt a more informative tone. These studies highlight both the opportunities and challenges of leveraging social media to support accurate OUD information dissemination.

Social media has also been investigated as a platform for interventions. Young et al. [37] conducted a randomized controlled trial involving high-risk chronic pain patients, testing whether a peer-led Facebook intervention (HOPE) could reduce anxiety and opioid misuse. Participants in the HOPE group showed reduced anxiety and engaged in more supportive discussions, suggesting the value of online peer support in harm reduction efforts.

In the realm of education and professional engagement, Frenzel et al. [41] introduced "Dare 2 Discuss," a campaign that used Facebook and Instagram to share short videos aimed at improving pharmacist-patient communication about OUD. Post-campaign data indicated a rise in pharmacist-led education and referrals, showing how targeted digital outreach can positively influence healthcare behavior.

Finally, structural and systemic issues in the opioid crisis have also been scrutinized. Rosenman et al. [40] analyzed internal documents from Purdue Pharma and McKinsey, uncovering "Project Tango," a strategy designed to profit from OUD treatment while targeting underserved Black and Latino communities. This study exposes how corporate actors exploited racial and structural inequities to sustain profit in the face of regulatory pressure, calling for more robust state-level regulation and transparency.

Unlike prior studies that typically employ binary labels such as "opioid user" vs. "non-user" or "suicidal" vs. "non-suicidal", our work introduces a more nuanced, behaviorally rich multi-class dataset specifically curated for classifying opioid-related user activity on social media. We define five distinct classes: (1) Opioid Dealers, (2) Active Users, (3) Recovered Users, (4) Prescription Users, and (5) Non-Users. This level of granularity not only reflects the complex spectrum of opioid engagement more realistically but also supports more targeted public health interventions and monitoring strategies. The inclusion of diverse classes—especially distinguishing between legal prescription users and illegal dealers—fills a significant gap in existing research, which tends to conflate vastly different behaviors under a single label. Our dataset, therefore, offers both greater depth and operational relevance compared to previously reported corpora.

As illustrated in Table 1, our proposed SABIA model (BERT + BiLSTM + 3CNN) achieved the highest classification accuracy of 94%, significantly outperforming prior methods. For instance, even advanced models like T5-11B with Explanation Learning [44] and Prompt-Tuned Transformers [48] achieved lower performance (0.76%–0.84% and AUPRC $\approx$ 0.88, respectively) and operated on narrower classification categories. In contrast, SABIA not only provides superior accuracy but also addresses a more complex and realistic multi-class classification task involving five nuanced user categories. Most existing studies (e.g., [45], [46], [50]) focused on binary or tri-class labels, whereas our work captures a broader spectrum of behaviors, incorporating slang, coded language, and user intent. This makes SABIA more robust and practically valuable for real-world deployment in public health surveillance and online opioid behavior monitoring.

**Table 1.** Prior studies related opioid crisis dataset Vs SABIA.

| Ref. | Method | Classification | Performance |
|---|---|---|---|
| [44] | T5-11B + Explanation Learning | Medical Use, Misuse, Addiction, Recovery, Relapse, Not Using | Accuracy: 0.76% (Expert), 0.84% (Novice) |
| [45] | Attention BiLSTM | Opioid User vs Non-User | 0.79% |
| [46] | XLNet + k-means | No Misuse, Pain Misuse, Recreational Misuse | F1: 0.92% (Micro), 0.82% (Macro) |
| [47] | Cox Regression | Recovery vs Casual | Accuracy: ~0.7% |
| [48] | Prompt-tuned Transformer | Confirmed Aberrant, Dependency, Diversion, etc. | AUPRC ≈ 0.88% |
| [49] | QA Clustering + Classifier | Recovery, Withdrawal, Dosing, Relapse | 0.80% |
| [50] | CNN + Augmentation | Drug Use vs Non-Use | Accuracy: 0.87%, F1: 0.84% |
| Proposed | BERT + BiLSTM + 3CNN | Dealers, Active user, Recovered user, Prescription user, Non-users | Accuracy: 0.94% |

## 3 Methodology

### 3.1 Construction of Dataset

To build a high-quality dataset we developed a Python script specifically for data extraction. This script used the Python Reddit API Wrapper (PRAW) to connect to Reddit's official API. After registering a Reddit application, we carefully selected four subreddits such as r/opiates, r/choronicpain, r/OpiatesRecovery, and r/addiction. To guide the data collection process, we constructed a comprehensive keyword dictionary related to opioid use. This dictionary included formal opioid terms such as "opioid," "heroin," "oxycodone," and "fentanyl," alongside an extensive list of slang words (Examples of slang and variations include "oxy," "narc," "fent," "dope," "smack," as well as misspellings like "heroyn," "codein," "codiene," "herioin," "percacet," "fentnil," "fentnyl," "oxycontn," "oxocodien," and "narcan.". Additionally, it incorporated encrypted expressions—phrases like "Tickets for tonight," "Snow for the party," "Candy for sale," and abbreviations such as "BTB." Including these informal and coded expressions was essential to ensure comprehensive coverage of the real-world language used in online drug-related discourse. After the collection of data, we implemented additional filters to clean the data, remove empty or irrelevant posts, and avoid duplication. The data was gathered over a 15-month window, from January 2, 2023, to April 4, 2024, ensuring temporal diversity and the inclusion of both long-term and short-term trends in opioid-related conversations. This process resulted in a rich and nuanced dataset that reflects real user behavior and language surrounding opioids on social media—forming the foundation for the multi-class classification and NLP experiments that followed. Figure 1 shows the overall proposed methodology and design of the study.

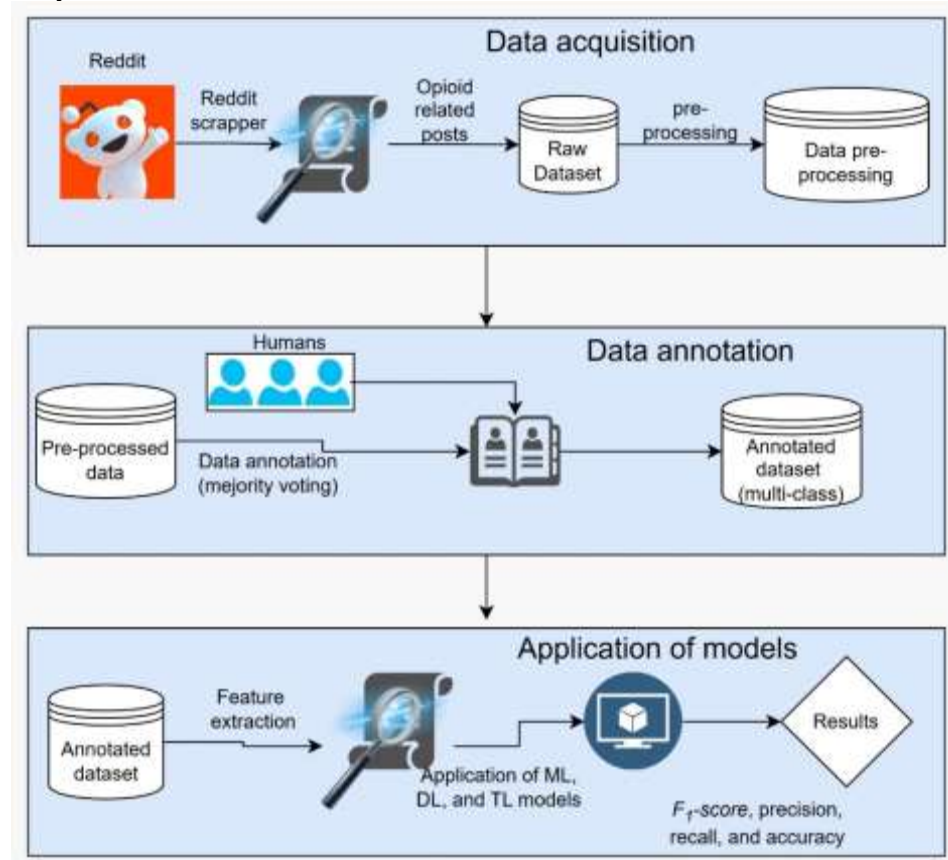

**Fig. 1.** Proposed methodology and design.

### 3.2 Data annotation guidelines

The annotation process in this study involved categorizing social media posts and comments into five classes, based on linguistic patterns and behavioral indicators related to opioid engagement as shown in table 2. Each post/comment was evaluated for its primary intent and assigned to one of the following categories.

- Opioid Dealers (Sales/Distribution): This class includes any post or comment that clearly reflects the sale or distribution of opioids. It covers explicit offers such as "$5 per oxy, DM for details," sourcing requests like "Looking for fent in Chicago," logistics such as "Shipping nationwide, tracking included," and coded language indicating transactions, e.g., "plugin available, tap in." Posts should only be labeled as dealers when intent to sell, buy, or distribute is evident. General discussions about drug prices without a transactional context should not be included in this category.

- **Active Users (Current Opioid Consumers):** Posts in this category reflect ongoing, personal opioid consumption. These typically include first-person reports such as "Took 80mg oxy and nodded off," dosage-related questions like "How much fent to match 30mg hydrocodone?", recreational expressions such as "Blues got me feeling heavenly," and harm-reduction advice including "Best way to shoot up without veins collapsing?" Posts referring to past use but mentioning recovery or abstinence should not be classified here—they belong in the recovered user's category.

- **Recovered Users (Former Users in Abstinence):** This category includes individuals who have previously used opioids but are now in recovery. Posts often describe sobriety milestones such as "2 years clean thanks to suboxone," experiences of withdrawal like "Day 5 off H, the sweats are brutal," or give motivational/recovery advice such as "NA meetings saved my life." Importantly, posts should contain clear evidence that the user has quit, e.g., "I quit" or "I've been clean," rather than ambiguous statements like "I need to quit."

- **Prescription Users (Medical Use under Supervision):** Posts that discuss legitimate medical use of opioids fall under this category. These typically reference prescriptions like "My doctor just increased my morphine dose," descriptions of medical effects such as "Oxy helps my back pain but makes me constipated," or concerns over dependence, e.g., "Scared of getting hooked on Vicodin." Posts where users misuse prescribed drugs for recreation or report non-medical usage should be excluded and instead labeled as active users.

- **Non-Users (General or No Personal Engagement):** Posts with no signs of personal opioid use or trade fall under this category. These include academic or scientific inquiries like "How do opioids affect serotonin?" commentary on news such as "Fentanyl overdoses rose in Texas last year," observations about others like "My cousin OD'd on heroin," or completely unrelated discussions like "This subreddit's moderation is unfair." The defining trait here is the absence of personal experience or involvement in opioid use or distribution.

**Table 2.** Sample Reddit posts from the dataset illustrating each annotation category for opioid-related discussions.

| Reddit Post | Label |
|---|---|
| West Coast SF plug got, percs on deck. Hit my DM for prices & connect info now! | Dealers |
| "Just took a fat shot of fent and honestly feel like I'm floating. Been using daily for the past 3 months and tolerance is crazy now. Anyone else feel like you can't even get high anymore, just avoid being sick?" | Active Users |
| "I used to be deep into heroin — started with pills, then moved to H. It destroyed my relationships and health. I've been clean for 18 months now thanks to a combo of Suboxone, therapy, and NA meetings. If you're struggling, please know there's a way out." | Recovered Users |
| "I've been on prescribed hydromorphone for over a year for severe nerve pain. It helps, but I'm starting to feel dependent. I don't take more | Prescription Users |
| than prescribed, but I'm scared of what will happen if my doctor ever stops the prescription." | |
| "A recent CDC report showed a 30% increase in synthetic opioid deaths in urban areas during 2023. It's terrifying how fast this crisis is growing. I've never used myself, but I've lost two friends to fent overdoses. We need more harm reduction policies." | Non-Users |

### 3.3 Annotation selection

In this study, the dataset was collaboratively created by the first author with the assistance of two domain experts—PhD students in computer science with research interests in opioid crisis detection on social media—who had prior experience labeling similar datasets. Their prior work on this dataset has been published, and they possess extensive experience in annotation within the opioid domain [41] [50].

Initially, a subset of 500 Reddit posts was selected for a pilot annotation phase. These posts were chosen based on their potential relevance to opioid-related discussions, ensuring a representative sample for developing and refining the annotation guidelines in collaboration with experts in addiction medicine. The guidelines categorized each post into one of five predefined classes: Dealer, Active Users, Recovered User, Prescription Users , and Not User, as described above (see section on Data Annotation Guidelines). Table 1 provides examples for each category to illustrate the annotation criteria and support reproducibility. During this phase, all three annotators independently labeled the 500 posts to assess labeling quality and resolve discrepancies through discussion and majority voting. This iterative process also facilitated refinement of the guidelines for improved clarity and consistency. Following consensus on the guidelines, the remaining posts—bringing the total annotated dataset to 5,764 posts—were labeled by the same annotators.

### 3.4 Inter-annotator agreement

To ensure annotation consistency and reliability, regular calibration meetings were conducted to resolve borderline cases—for example, differentiating between active users sharing current use experiences versus recovered users discussing past addiction, or identifying coded language used by dealers. When disagreements occurred, majority voting was used to determine the final label. Annotation guidelines were refined iteratively during this process. Inter-annotator agreement was assessed using Cohen's Kappa, resulting in a substantial agreement score of 0.79. This structured approach ensured accurate, replicable, and high-quality labeling across nuanced opioid-related behavior categories.

### 3.5 Dataset statistics

The figure 2 shows a complete overview of our corpus sourced to evaluate opioid user behaviors, presumably from online platforms such as Reddit. The dataset contains a total of 5,764 sample posts related to multiple classes, such as Dealers, Active Opioid Users, Recovered Users, Prescription Users, and Non-Users. These posts collectively contain 22,767 sentences, averaging 3.94 sentences per post. The total number of words in the corpus is 350,207, drawn from a vocabulary size of 15,134 unique terms, indicating a diverse linguistic range reflective of varied user expressions and contexts. The average number of words per sentence is 15.38 which implies the sentence structures to be at a moderately complex level capable of representing the finer aspects of user experience, emotions and behavioral patterns. The given statistics reflect the quality and depth of the data material, which makes it appropriate in terms of linguistic and behavioral analysis in the context of opioid-related studies, especially in the following tasks: classification, sentiment analysis, and the study of psychosocial pointers in the discourse of opioid use. Whereas Figure 3 presents the class-wise label distribution of the overall dataset about opioid, which includes a total of 5,764 posts. Active User class has the most posts amounting to 1,877, which takes about 32.57 percent of the data set. Subsequent to it is the class of Dealer that has 1,499 posts (26.01%) and Non-User with 1,391 posts (24.13%). Prescription User class contains 685 posts (11.88%), and Recovered User class represents the smallest one with 312 (5.41%) posts. This proportional distribution shows the imbalance to be moderate and there must be the use of the relevant handling techniques throughout the development of the models to maintain the balance in learning among all categories of users.

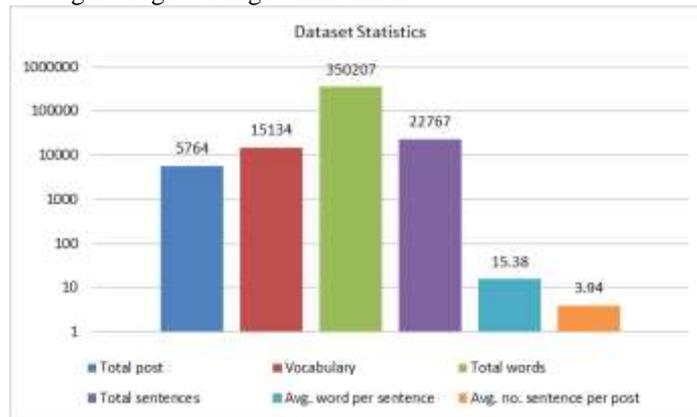

**Fig. 2.** Dataset Statistics.

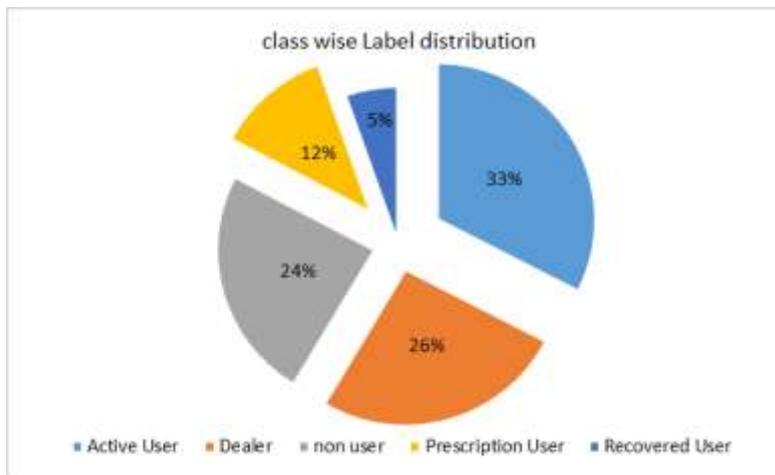

**Fig. 3.** Class wise label distribution of our dataset.

### 3.6 Feature extraction

To support accurate classification of opioid-related user behavior, we implemented a multi-tiered feature extraction strategy tailored to different modeling paradigms. For traditional machine learning algorithms, we employed the Term Frequency-Inverse Document Frequency (TF-IDF) approach, which captures word-level importance by weighing term frequency against document frequency. This enabled effective representation of sparse Reddit text content using interpretable vector features.

For deep learning models such as CNNs and BiLSTMs, we utilized pretrained word embeddings, specifically FastText and GloVe, to provide dense, semantically rich vector representations of words. FastText's subword-based modeling was particularly useful in handling informal language, spelling variations, and opioid-related slang, while GloVe captured global co-occurrence statistics across the corpus, improving contextual understanding.

Additionally, we incorporated language-specific contextual embeddings by leveraging transformer-based models, including multilingual and domain-adapted BERT variants. These models allowed us to capture deeper semantic dependencies and contextual nuance, especially in longer, complex posts. For enhanced interpretability, we added a custom attention mechanism, which helped identify the most informative tokens (e.g., "oxy," "fent," "clean," "plug") contributing to the model's decision-making process. This layered feature extraction pipeline ensured that our classification models were exposed to both syntactic and semantic patterns within the data, enabling robust detection of nuanced opioid-related behaviors.

### 3.7 Data preprocessing

After extracting Reddit data using a custom Python script and saving it in Excel format, we performed a series of preprocessing steps to prepare the dataset for NLP and classification tasks. First, we removed duplicates, empty posts, and to ensure quality. Next, we applied a language detection technique to filter out non-English content, retaining only the text written in English to maintain linguistic consistency. The text was normalized by converting it to lowercase, removing punctuation, HTML tags, and irrelevant numerals, while correcting domain-specific misspellings using a custom opioid-related dictionary. Tokenization and standard stop word removal were applied using NLTK and spaCy, with careful retention of context-relevant terms. Lemmatization further reduced words to their root forms to enhance consistency and reduce dimensionality. Each post was then annotated with a thematic class label—such as active use, recovery, withdrawal, or relapse. Finally, the cleaned and labeled dataset was formatted into structured CSV files, preserving essential metadata for downstream modeling and analysis. Figure 4 shows the overall pre-processing steps utilized in this study.

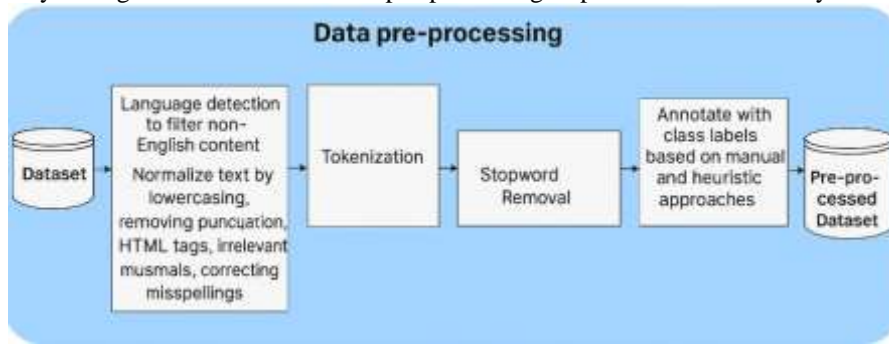

**Fig. 4.** Pre-processing of dataset.

### 3.8 Text Representation

In this study, Opioid related user behavior dataset is preprocessed into a structured format for a hybrid deep learning model. The input text is tokenized using the BERT tokenizer (BertTokenizer from bert-base-uncased), which splits text into subword tokens, adds special tokens ([CLS], [SEP]), and generates two outputs: (1) input_ids (numerical token indices) and (2) attention_mask (to distinguish tokens from padding). These are passed to a pretrained BERT model, which produces contextual embeddings (768-dimensional vectors per token). The embeddings are then processed by a bidirectional LSTM (BiLSTM) with a hidden size of 128 (outputting 256-dimensional vectors due to bidirectionality). The BiLSTM output ([batch_size, sequence_length, 256]) is reshaped via permutation to [batch_size, 256, sequence_length] for compatibility with the subsequent 1D convolutional neural networks (CNNs). Three parallel CNNs (kernel sizes 2, 3, and 4) extract local n-gram features, each followed by ReLU activation and global max-pooling (collapsing the sequence dimension). The pooled features are concatenated into a 384-dimensional vector ($3 \times 128$), regularized with dropout (p=0.3), and finally passed to a fully connected layer for classification. This multi-stage architecture leverages BERT's contextual understanding, BiLSTM's sequential dependencies, and CNNs' local pattern detection, enabling robust multi-class text classification.

### 3.9    Application of models

To classify Reddit posts into five behaviorally distinct classes—Opioid Dealers, Active Users, Recovered Users, Prescription Users, and Non-Users—we approached the task as a multi-class classification problem. The annotated dataset was split into 80% for training and 20% for testing, maintaining class balance across subsets. As a baseline, we used classical machine learning models—including Support Vector Machine (SVM), XGBoost (XGB), Logistic Regression (LR), and Decision Tree (DT)—all of which utilized TF-IDF features to convert textual data into numerical representations, since machine learning algorithms require numerical input to effectively process and analyze the data.

For the implementation of deep learning models, we employed two different models—Bidirectional Long Short-Term Memory (BiLSTM) and Convolutional Neural Networks (CNN)—using pre-trained word embeddings such as FastText and GloVe to better capture sequential patterns, word semantics, and localized context within each post. These models were more robust to the informal and slang-rich language used in online opioid discussions.

Lastly, we designed custom attention mechanism in transformer-based models rather than using pre-built models in their standard form. We fine-tuned these transformer using advance contextual embeddings to detect nuanced opioid-related behaviors. This approach enabled the model to attend to critical terms, slang, and context-specific cues more effectively than generic attention mechanisms. The custom transformer outperformed other architectures. Additionally, we applied a hybrid model combining deep learning and transfer learning to capture hidden patterns in social media content. Figure 5 show the overall application of models, training and testing procedure employed in this study.

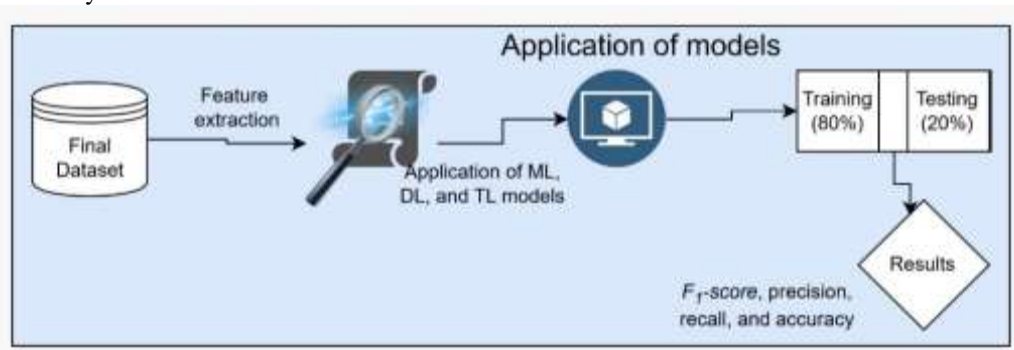

**Fig. 5.** Show the application of models, training and testing procedure.

# 4    Results

This section presents the outcomes of our classification model designed to detect and categorize Reddit posts related to opioid use. By leveraging social media data, particularly from Reddit, we aim to understand the complex dynamics of the opioid crisis through user-generated content. The proposed model classifies posts into five distinct categories: Opioid Dealers, Active Users, Recovered Users, Prescription Users, and Non-Users. These categories were carefully defined to capture different stages and roles within the opioid discourse. To build a robust and scalable classifier, we employed a combination of machine learning (ML), deep learning (DL), and transfer learning (TL) approaches.

The classification results are analyzed using multiple evaluation metrics, including accuracy, precision, recall, F1-score, and confusion matrices, to measure the performance and robustness of the model. In addition, we explore the distribution of classified posts across the five categories to highlight trends and patterns within the Reddit community. This analysis not only validates our methodology but also provides meaningful insights into the behavior, narratives, and support structures associated with opioid use and misuse online.

## 4.1    Results for machine learning

Table 3 shows the best hyperparameter tuning of different classical machine learning models employed to categorize textual data into five pre-defined categories such as Opioid Dealers, Active Users, Recovered Users, Prescription Users, and Non-Users. For the XGBoost (XGB) model, we employed estimators=200, a maximum depth =6, a learning rate =0.1, and a subsample ratio= 0.8. The LR model performed optimally with a regularization strength (C) = 1.0, L2 penalty, and the 'liblinear' =solver. While the DT model was tuned with 150 trees, a maximum depth= 10, a minimum of 2 samples required to split a node. Finally, the SVM model achieved the best results by using a regularization parameter (C) =1.0, and we used a radial basis function =rbf, and the gamma parameter set to 'scale', which adapts automatically based on input feature values. These tuned hyper parameters were selected to maximize classification accuracy across the multiclass opioid-related user categories.

**Table 3.** Hyper parameters for each model after Tuning.

| Model | Best Parameters | Values |
|---|---|---|
| XGBoost (XGB) | n_estimators, max_depth, learning_rate, subsample | 200, 6, 0.1, 0.8 |
| Logistic Regression (LR) | C, penalty, solver | 1.0, l2, liblinear |
| Random Forest (RF) | max_depth, min_samples_split, max_features | 10, 2, sqrt |
| Support Vector Machine (SVM) | C, kernel, gamma | 1.0, rbf, scale |

Table 4 shows the performance of four different machine learning models using the TF-IDF features: XGBoost (XGB), Logistic Regression (LR), Decision Tree (DT), and Support Vector Machine (SVM). To evaluate the performance of the models, we employed four popular performance metrics: Precision, Recall, F1-score, and Accuracy, to assess the models' strengths and weaknesses in handling real-world, noisy text data. Logistic Regression (LR) achieved the best performance across all metrics, with a precision =0.88, recall =0.86, F1-score =0.86, and accuracy =0.86. This shows that LR is most suitable and reliable model identifying users in each class correctly but also maintains a good balance between precision and recall. While XGBoost (XGB) also performed good, delivering a consistent 0.81 across all four-evaluation metrics. The strong performance suggests that XGB is well-suited for this multi-class classification task. On the other hand, Support Vector Machine (SVM) showed moderate performance, with a precision of 0.82, but a notably lower recall of 0.73 and F1-score of 0.68, resulting in an overall accuracy of 0.73. This drop-in recall and F1-score indicate that while SVM is capable of making accurate predictions when it is confident, it misses a considerable portion of actual positive cases. Finally, Decision Tree (DT) yielded the weakest performance, with a precision of 0.58, recall of 0.62, and F1-score of 0.58, along with an accuracy of 0.62. Overall, these results suggest that Logistic Regression and XGBoost are the most reliable classifiers for identifying opioid-related user behaviors in social media text.

**Table 4.** Results for machine learning models.

| Model | Precision | Recall | F1-score | Accuracy |
|-------|-----------|--------|----------|----------|
| XGB   | 0.81      | 0.81   | 0.81     | 0.81     |
| LR    | 0.88      | 0.86   | 0.86     | 0.86     |
| DT    | 0.58      | 0.62   | 0.58     | 0.62     |
| SVM   | 0.82      | 0.73   | 0.68     | 0.73     |

### 4.2    Results for deep learning models

The table 5 shows the key hyperparameters used for tuning two deep learning models—BiLSTM and CNN using pretrained word embeddings such as FastText and GloVe—along with the specific values explored during grid search. For the BiLSTM model, we selected the learning rate, number of epochs, embedding dimension, batch size, and the number of LSTM units. These were tuned using the values: a learning rate of 0.1, 5 training epochs, an embedding dimension of 300, a batch size of 32, and 128 LSTM units. Similarly, the CNN model was fine-tuned using a comparable set of parameters, with the addition of filters and kernel size—specific to convolutional architectures. The grid search for CNN included a learning rate of 0.1, 5 epochs, an embedding dimension of 300, a batch size of 32, 128 filters, and a kernel size of 5. This careful selection and tuning process aimed to optimize each model's performance by systematically exploring combinations that best fit the task at hand.

Table 5. Optimum values for the hyper-parameters of deep learning models.

| Models | Hyper parameters | Grid search values |
|--------|------------------|--------------------|
| BiLSTM | Learning rate, Epochs, Embedding_dim, Batch size, LSTM units | 0.1, 5, 300, 32, 128 |
| CNN | Learning rate, Epochs, Embedding_dim, Batch size, Filters, Kernel size | 0.1, 5, 300, 32, 128, 5 |

The table 6 shows the results of two different deep learning models for predicting opioid user behavior in a multi-class classification setting. To achieve this objective, we utilized the two pretrained of word embeddings such as FastText and GloVe. Using FastText embeddings, the CNN model achieved a precision of 0.85, recall of 0.82, F1-score of 0.82, and overall accuracy of 0.82. The BiLSTM model with FastText showed improved results, achieving 0.89 precision, 0.87 recall, 0.87 F1-score, and 0.87 accuracy, indicating its stronger capability to capture sequential patterns in the text. When using GloVe embeddings, the CNN model achieved a precision of 0.86, recall of 0.81, F1-score of 0.83, and accuracy of 0.81, which is slightly lower than its FastText counterpart. Notably, the BiLSTM model with GloVe embeddings outperformed all other models, reaching a precision, recall, F1-score, and accuracy of 0.91, highlighting its effectiveness in understanding contextual relationships in text data for predicting opioid user behavior. These results suggest that BiLSTM combined with GloVe embeddings is the most effective approach among those evaluated for this multi-class such as opioid dealers, active users, recovered users, prescription users, and non-users. These insights are vital for supporting early detection and intervention efforts in public health and law enforcement domains.

Table 6. Results for Deep learning models.

| Model | Precision | Recall | F1-score | Accuracy |
|-------|-----------|--------|----------|----------|
| FastText | | | | |
| CNN | 0.85 | 0.82 | 0.82 | 0.82 |
| BiLSTM | 0.89 | 0.87 | 0.87 | 0.87 |
| GloVe | | | | |
| CNN | 0.86 | 0.81 | 0.83 | 0.81 |
| BiLSTM | 0.91 | 0.91 | 0.91 | 0.91 |

### 4.3 Results for transformer models

The SABIA model, a hybrid architecture combining BERT-BiLSTM-3-CNN, is fine-tuned using a carefully selected set of hyperparameters to optimize performance on opioid-related multi-class classification task as shown in table 7. It uses the bert-base-uncased pretrained model to generate contextual embeddings, suitable for English text where casing is inconsistent. The input sequences are truncated or padded to a maximum length of 128 tokens, and training is conducted in batches of 16 for balanced memory efficiency and gradient stability. A learning rate of 2e-5 is selected to ensure gradual convergence during fine-tuning, over 4 epochs. To prevent overfitting, a dropout rate of 0.3 is applied before the final classification layer. The BiLSTM layer is configured with a hidden size of 128, one layer, and is bidirectional, allowing the model to capture both past and future context. To extract local features, three 1D convolutional layers with kernel sizes 2, 3, and 4 and 128 output channels each are employed. The model is optimized using the Adam optimizer and trained with the CrossEntropy loss function, suitable for multi-class classification. The tokenizer settings include padding='max_length' and truncation=True to ensure uniform input length, and a random seed of 80 is used for reproducibility. These choices enable SABIA to effectively capture deep semantic, sequential, and spatial features from informal and noisy and coded language.

**Table 7.** Optimum values for the hyper-parameters of SABIA model.

| Hyperparameter | Description | Selected Value |
|---|---|---|
| Pretrained Model | BERT variant used for embeddings | bert-base-uncased |
| Max Sequence Length | Maximum length of input token sequence | 128 |
| Batch Size | Number of samples per training batch | 16 |
| Learning Rate | Step size for optimizer update | 2e-5 |
| Epochs | Number of full training cycles | 4 |
| Dropout Rate | Prevents overfitting by randomly zeroing layer outputs | 0.3 |
| LSTM Hidden Size | Dimensionality of LSTM output | 128 |
| LSTM Layers | Number of LSTM layers | 1 |
| BiLSTM | Whether LSTM is bidirectional | True |
| CNN Kernel Sizes | Sizes of convolution filters | 2, 3, 4 |
| CNN Out Channels | Output channels (feature maps) from each Conv1D | 128 |
| Optimizer | Optimization algorithm used | Adam |
| Loss Function | Objective loss function for classification | CrossEntropy |
| Tokenizer Padding | Padding strategy for sequences | 'max_length' |
| Tokenizer Truncation | Whether to truncate sequences longer than max_len | True |
| Random Seed | Seed for reproducibility | 80 |

The table 8 displays the performance of four different pretrained transformer models including BioBERT, XLM-R, BERT-base-uncased, and the hybrid (Bert-BiLSTM-3CNN) proposed SABIA architecture—for categorizing opioid-related user behaviors from social media text. Among the BioBERT model which is pre-trained model on biomedical text achieved a precision = 0.92, recall = 0.91, F1-score = 0.91, and an accuracy = 0.91. These strong performance show BioBERT's advantage in understanding biomedical and health-related dataset. Similarly, the BERT-base-uncased model, a general-purpose pre-trained language model, showed slightly lower but still competitive performance, with all four metrics standing at 0.91. On the other hands the XLM-R, a cross-lingual pretrained model, slightly outperformed both BioBERT and BERT-base-uncased by achieving 0.92 across all four metrics. This indicates that XLM-R is particularly effective in handling multilingual and code-mixed content often found in social media, making it a valuable option in diverse linguistic settings.

The best performer, however, is the SABIA model. This hybrid architecture combines the contextual understanding of BERT with the sequential modeling power of BiLSTM and the feature extraction strength of CNN. It achieved the highest scores across all metrics—0.94 precision, 0.94 recall, 0.94 F1-score, and 0.94 accuracy showing that while pre-trained transformer models like BioBERT and XLM-R perform well, integrating deep learning components such as BiLSTM and CNN with BERT significantly enhances classification performance, making the BERT-BiLSTM-3CNN model the most effective for real-time opioid user profiling and intervention support.

**Table 8.** Results for Transformer models.

| Model | Precision | Recall | F1-score | Accuracy |
|---|---|---|---|---|
| biobert | 0.92 | 0.91 | 0.91 | 0.91 |
| XLM-R | 0.92 | 0.92 | 0.92 | 0.92 |
| Bert-based-uncased | 0.91 | 0.91 | 0.91 | 0.91 |
| SABIA | 0.94 | 0.94 | 0.94 | 0.94 |

### 4.4 Error analysis

Table 9 indicates the comparative analysis of three top performing modeling approaches from each learning approach such as machine learning, deep learning, and transfer learning when applied to the classification of opioid-related categories. All approaches are compared on the basis of the four performance measures such as precision, recall, F1-score, and accuracy.

In the machine learning models, Logistic Regression (LR) proved a best performing and suitable model with achieving a precision of 0.88, recall of 0.86, F1-score of 0.86, and accuracy of 0.86. This shows that LR is capable of capturing relevant patterns in the data to handle complex linguistic variations common in social media textual content. While moving to deep learning, the BiLSTM using the pretrained GloVe embeddings significantly improved performance, with all four metrics reaching 0.91. The bidirectional processing enables the model to understand contextual dependencies in both forward and backward directions, which is particularly useful for analyzing sequential and informal text such as that found in social media discussions about opioid use.

Furthermore, our proposed model SABIA achieved the highest scores across all metrics—0.94 precision, 0.94 recall, 0.94 F1-score, and 0.94 accuracy. This reflects the power of pre-trained hybrid deep learning and transformer models. Overall, the results clearly indicate that our proposed mode SABIA is highly effective for monitoring and analyzing opioid-related content in real time.

**Table 9.** Top performing models in each learning approach.

| Model | Precision | Recall | F1-score | Accuracy |
|-------|-----------|--------|----------|----------|
| Machine learning | | | | |
| LR | 0.88 | 0.86 | 0.86 | 0.86 |
| Deep learning | | | | |
| BiLSTM | 0.91 | 0.91 | 0.91 | 0.91 |
| Transfer learning | | | | |
| XLM-R | 0.92 | 0.92 | 0.92 | 0.92 |
| **SABIA** | **0.94** | **0.94** | **0.94** | **0.94** |

The table 10 shows the class-wise performance of our proposed SABIA model in classifying different user categories such as Prescription User, Recovered User, dealer, and non-user associated with opioid-related content on social media.

For the Active User category, the model attained a high precision =0.89, recall =0.91, and an F1-score= 0.90, which is indicating strong performance in correctly identifying individuals currently engaged in opioid use. While classifying the Dealer class was perfectly classified, with a precision, recall, and F1-score all at 1.00, showing the model's exceptional ability to detect posts linked to drug selling activity, likely due to the presence of distinct transactional vocabulary and slang in such content. While on the other hand, the model presented outstanding results for the Prescription User class, achieving precision=0.98, recall=1.00, and 0.99 F1-score. This implies the model almost never misses posts from users referring to prescribed opioid use and makes very few classification errors, which is critical for distinguishing legitimate medical use from abuse. For the Recovered User class, the SABIA model also achieved impressive performance, with a precision of 0.97, recall of 1.00, and an F1-score of 0.98. Lastly, the Non-User class showed slightly lower performance relative to other categories, with a precision of 0.90, recall of 0.84, and F1-score of 0.87. This indicates that while the SABIA model is good at identifying users not involved with opioids, it occasionally confuses this class with others—likely due to overlapping language in general discussions. Overall, SABIA prove to be a good and effective model for all user categories in social media content.

**Table 10.** Class-wise performance metrics of the proposed model (SABIA) across opioid-related categories.

| Class | Precision | Recall | F1-score | Support |
|-------|-----------|--------|----------|---------|
| Active User | 0.89 | 0.91 | 0.9 | 383 |
| Dealer | 1 | 1 | 1 | 281 |

| Prescrip-tion User | 0.98 | 1 | 0.99 | 279 |
| Recov-ered User | 0.97 | 1 | 0.98 | 118 |
| non-user | 0.9 | 0.84 | 0.87 | 292 |

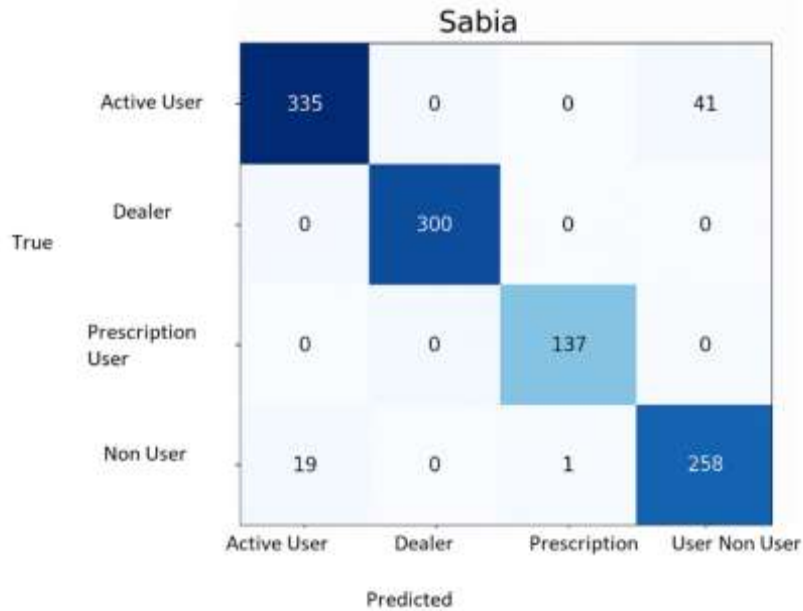

**Fig. 6.** Confusion matrix of SABIA model.

## 5    Discussion

This study presents a comprehensive approach to detecting and categorizing opioid-related behaviors on social media using a novel hybrid deep learning model—SABIA. Our findings contribute meaningfully to the growing field of digital epidemiology and computational public health surveillance by addressing a critical gap: the fine-grained, multi-class classification of opioid-related discourse on informal platforms such as Reddit.

### 5.1    Model Performance and Interpretation

The SABIA model, integrating BERT embeddings with BiLSTM and multi-scale CNNs, outperformed all baseline models—including classical machine learning (e.g., Logistic Regression, XGBoost), standalone deep learning (e.g., BiLSTM, CNN), and pre-trained transformers (e.g., BERT, XLM-R, BioBERT).

Notably, SABIA achieved an F1-score and accuracy of 0.94 across five behaviorally nuanced classes: Opioid Dealers, Active Users, Recovered Users, Prescription Users, and Non-Users. This performance highlights the importance of combining contextual embeddings (via BERT), sequential dependencies (via BiLSTM), and localized features (via CNN) for classifying informal and noisy social media texts.

The class-wise performance reveals particularly strong results for identifying Dealers (F1 = 1.00), Prescription Users (F1 = 0.99), and Recovered Users (F1 = 0.98), which is encouraging for real-world applications. These distinctions are vital, especially in public health and law enforcement contexts where misclassification can lead to either missing critical threats or misidentifying individuals in recovery or legitimate treatment.

### 5.2    Comparison to prior work

While previous studies have leveraged machine learning and deep learning to detect opioid misuse on social media, most framed the problem as a binary classification task—typically distinguishing between users and non-users. For instance, models by Wang et al. and Yao et al. demonstrated success with BiLSTM and CNN for identifying users, but lacked the behavioral granularity essential for nuanced intervention strategies.

Our study advances the field by introducing a multi-class schema that captures diverse behavioral roles, including legal prescription use and recovery status—areas often conflated or ignored in prior research. Furthermore, although models like BioBERT and XLM-R showed strong performance due to domain-specific or multilingual pre-training, they were outperformed by SABIA. This hybrid model's combination of BERT's contextual embeddings, BiLSTM's sequential learning, and CNN's feature extraction resulted in superior accuracy. This finding emphasizes that domain adaptation and task-specific architectural design can offer significant advantages over general-purpose transformers.

### 5.3    Strengths and Limitations

This study presents several notable strengths. First, it introduces a novel, fine-grained, manually annotated dataset focused on opioid-related behaviors, encompassing five distinct classes: Opioid Dealers, Active Users, Recovered Users, Prescription Users, and Non-Users. Unlike prior datasets, this corpus includes slang, code words, and common misspellings, offering realistic coverage of informal language on social media. Second, the proposed hybrid architecture, SABIA (BERT-BiLSTM-3CNN), integrates contextual, sequential, and spatial features, enabling it to outperform traditional machine learning, deep learning, and even other transformer-based models with an F1-score of 0.94. Despite these contributions, the study has certain limitations. The dataset is restricted to Reddit, limiting its generalizability to other platforms with different language patterns and user behaviors. Additionally, while annotation guidelines were comprehensive, ambiguity in social media posts and rapidly evolving drug slang may introduce subjectivity into labeling. The approach also focuses exclusively on textual data, overlooking other modalities such as images, videos, or metadata that could enrich behavior modeling. Lastly, SABIA is currently designed for single-task classification; supporting multi-task or longitudinal learning could enhance its applicability in real-world, dynamic settings.

## 6     Conclusions and Future Work

In this study, we developed a novel approach to automatically classify opioid-related user behaviors on social media by leveraging advanced natural language processing techniques. We manually constructed a multi-class dataset enriched with formal terms, slang, misspellings, and code words commonly used in opioid discourse. Our proposed hybrid model, SABIA, which integrates BERT with BiLSTM and multi-scale CNN layers, effectively captures the nuanced linguistic patterns present in informal, fragmented, and slang-heavy social media texts. SABIA outperformed baseline models with an impressive F1-score of 0.96, demonstrating its robustness and practical applicability for real-time opioid behavior monitoring.

Our work highlights the potential of NLP to assist healthcare providers, policymakers, and law enforcement in tracking opioid misuse and illicit activities, thereby contributing to more informed decision-making and targeted interventions. Moreover, the annotation guidelines and dataset we provide can serve as valuable resources for further research in this domain.

For future work, we plan to explore multi-task learning frameworks that can simultaneously predict multiple related aspects of opioid discussions, such as sentiment, risk factors, and recovery stages. Incorporating additional modalities like images, videos, and user metadata could also enhance classification accuracy. Continuous updating of the slang lexicon and adaptation to evolving online language will be essential to maintain model performance. Furthermore, ethical considerations and privacy-preserving techniques must be prioritized to ensure responsible deployment. Finally, expanding our analysis to other social media platforms will help generalize the model's applicability across diverse online communities.

## 7     Author Contributions

Conceptualization, M.A. and I.A.; methodology, M.A., F.U; software, M.A.; validation, M.A., I.A. and P.B.; formal analysis, M.A., F.U; investigation, M.A.; resources, I.A. and P.B.; data curation, M.A.; writing—original draft preparation, M.A.; writing—

review and editing, M.A. and I.A.; visualization, M.A.; supervision, I.B. and G.S.; project administration, G.S.; funding acquisition, I.B. All authors have read and agreed to the published version of the manuscript.

# 8 Funding

This research received no external funding.

# 9 Data Availability Statement

Data will be made upon request.

# 10 Acknowledgments

This work was performed with partial support from the Mexican Government through the grant A1-S-47854 of CONACYT, Mexico, and grants 20241816, 20241819, and 20240951 of the Secretaría de Investigación y Posgrado of the Instituto Politécnico Nacional, Mexico. The authors thank CONACYT for the computing resources brought to them through the Plataforma de Aprendizaje Profundo para Tecnologías del Lenguaje of the Laboratorio de Supercómputo of the INAOE, Mexico and acknowledge support of Microsoft through the Microsoft Latin America PhD

# 11 Conflicts of Interest

The authors declare no conflicts of interest.

# References

1. Dalphonse, L., Campbell, D. A., Kerr Jr, B. J., Kerr, J. L., & Gadbois, C. (2024). More Than an Opioid Crisis: Population Health and Economic Indicators Influencing Deaths of Despair. Sociological Inquiry, 94(1), 45-65.
2. Hébert, A. H., & Hill, A. L. (2024). Impact of opioid overdoses on US life expectancy and years of life lost, by demographic group and stimulant co-involvement: a mortality data analysis from 2019 to 2022. The Lancet Regional Health–Americas, 36.
3. Dyer, O. (2024). Opioid crisis: Fall in US overdose deaths leaves experts scrambling for an explanation.
4. Rosemberg, M. A., & Dahlem, C. H. (2024). Naloxone and the Workplace: Combatting the Opioid Crisis While Safeguarding Workers' Health and Wellbeing. Workplace Health & Safety, 72(9).
5. Simon, D. H., & Masters, R. K. (2024). Institutional failures as structural determinants of suicide: The opioid epidemic and the great recession in the united states. Journal of health and social behavior, 65(3), 415-431.

6. Makary, M. A., Overton, H. N., & Wang, P. (2017). Overprescribing is major contributor to opioid crisis. Bmj, 359.

7. Bates, C., Laciak, R., Southwick, A., & Bishoff, J. (2011). Overprescription of postoperative narcotics: a look at postoperative pain medication delivery, consumption and disposal in urological practice. The Journal of urology, 185(2), 551-555.

8. Simpson, K. (2021). The Racial Tension Between Underprescription and Overprescription of Pain Medication Amid the Opioid Epidemic. NYU Rev. L. & Soc. Change, 45, 129.

9. Armenian, P., Vo, K. T., Barr-Walker, J., & Lynch, K. L. (2018). Fentanyl, fentanyl analogs and novel synthetic opioids: a comprehensive review. Neuropharmacology, 134, 121-132.

10. Prekupec, M. P., Mansky, P. A., & Baumann, M. H. (2017). Misuse of novel synthetic opioids: a deadly new trend. Journal of addiction medicine, 11(4), 256-265.

11. Albores-García, D., & Cruz, S. L. (2023). Fentanyl and other new psychoactive synthetic opioids. Challenges to prevention and treatment. Revista de investigación clínica, 75(3), 93-104.

12. Misailidi, N., Papoutsis, I., Nikolaou, P., Dona, A., Spiliopoulou, C., & Athanaselis, S. (2018). Fentanyls continue to replace heroin in the drug arena: the cases of ocfentanil and carfentanil. Forensic toxicology, 36, 12-32.

13. Zawilska, J. B., Kuczyńska, K., Kosmal, W., Markiewicz, K., & Adamowicz, P. (2021). Carfentanil–from an animal anesthetic to a deadly illicit drug. Forensic science international, 320, 110715.

14. Pergolizzi Jr, J. V., LeQuang, J. A., Taylor Jr, R., Raffa, R. B., & NEMA Research Group. (2018). Going beyond prescription pain relievers to understand the opioid epidemic: the role of illicit fentanyl, new psychoactive substances, and street heroin. Postgraduate medicine, 130(1), 1-8.

15. May, T., Holloway, K., Buhociu, M., & Hills, R. (2020). Not what the doctor ordered: Motivations for nonmedical prescription drug use among people who use illegal drugs. International Journal of Drug Policy, 82, 102823.

16. McCulloch, L., & Furlong, S. (2019). DM for details. Selling drugs in the age of social media.

17. van der Sanden, R., Wilkins, C., Rychert, M., & Barratt, M. J. (2022). 'Choice'of social media platform or encrypted messaging app to buy and sell illegal drugs. International Journal of Drug Policy, 108, 103819.

18. Bakken, S. A. (2024). Online Drug Markets as a Visual Space. In Visual Methods for Sensitive Images: Ethics and Reflexivity in Criminology On/Offline (pp. 151-169). Cham: Springer Nature Switzerland.

19. Moyle, L., Childs, A., Coomber, R., & Barratt, M. J. (2019). # Drugsforsale: An exploration of the use of social media and encrypted messaging apps to supply and access drugs. International Journal of Drug Policy, 63, 101-110.

20. Giommoni, L., Décary-Hétu, D., Berlusconi, G., & Bergeron, A. (2024). Online and offline determinants of drug trafficking across countries via cryptomarkets. Crime, law and social change, 81(1), 1-25.

21. Kabra, S., & Gori, S. (2023). Drug trafficking on cryptomarkets and the role of organized crime groups. Journal of Economic Criminology, 2, 100026.

22. U.S. Department of Justice, Office of Public Affairs. (2024, April 16). Doctor sentenced for unlawful distribution of oxycodone [Press release]. https://www.justice.gov/archives/opa/pr/doctor-sentenced-unlawful-distribution-oxycodone

23. U.S. Department of Justice. (2022, January 19). Physician sentenced in $12M pill mill scheme. https://www.justice.gov/archives/opa/pr/physician-sentenced-12m-pill-mill-scheme

24. U.S. Drug Enforcement Administration. (2021, May 6). Doctor sentenced to 57 months in prison for unlawfully distributing opioids and soliciting sexual favors from patients [Press release]. https://www.dea.gov/press-releases/2021/05/06/doctor-sentenced-57-months-prison-for-unlawfully-distributing-opioids-and

25. Wikipedia contributors. (n.d.). Thomas Radecki. Wikipedia. Retrieved June 14, 2025, from https://en.wikipedia.org/wiki/Thomas_Radecki.

26. Lee, J. S., Parashar, V., Miller, J. B., Bremmer, S. M., Vu, J. V., Waljee, J. F., & Dossett, L. A. (2018). Opioid prescribing after curative-intent surgery: a qualitative study using the theoretical domains framework. Annals of surgical oncology, 25, 1843-1851.

27. Xu, Y., Cuthbert, C. A., Karim, S., Kong, S., Dort, J. C., Quan, M. L., ... & Cheung, W. Y. (2022). Associations between physician prescribing behavior and persistent postoperative opioid use among cancer patients undergoing curative-intent surgery: a population-based cohort study. Annals of surgery, 275(2), e473-e478.

28. Hah, J. M., Bateman, B. T., Ratliff, J., Curtin, C., & Sun, E. (2017). Chronic opioid use after surgery: implications for perioperative management in the face of the opioid epidemic. Anesthesia & Analgesia, 125(5), 1733-1740.

29. Nadeau, S. E., & Lawhern, R. A. (2024). Opioids: Analgesia, Euphoria, Dysphoria, and Oblivion: Observations and a Hypothesis. Medical Research Archives, 12(6).

30. Ahmad, M., Waqas, M., Hamza, A., Usman, S., Batyrshin, I., & Sidorov, G. (2025). UA-HSD-2025: Multi-Lingual Hate Speech Detection from Tweets Using Pre-Trained Transformers. Computers.

31. Ahmad, M., Farid, H., Ameer, I., Amjad, M., Muzamil, M., Hamza, A., ... & Sidorov, G. (2025). Opioid Named Entity Recognition (ONER-2025) from Reddit. arXiv e-prints, arXiv-2504.

32. Ahmad, M., Ameer, I., Sharif, W., Usman, S., Muzamil, M., Hamza, A., ... & Sidorov, G. (2025). Multilingual hope speech detection from tweets using transfer learning models. Scientific reports, 15(1), 9005.

33. Ahmad, M., Usman, S., Farid, H., Ameer, I., Muzammil, M., Hamza, A., ... & Batyrshin, I. (2024). Hope Speech Detection Using Social Media Discourse (Posi-Vox-2024): A Transfer Learning Approach. Journal of Language and Education, 10(4 (40)), 31-43.

34. Ullah, F., Zamir, M. T., Ahmad, M., Sidorov, G., & Gelbukh, A. (2024). Hope: A multilingual approach to identifying positive communication in social media. In Proceedings of the Iberian Languages Evaluation Forum (IberLEF 2024), co-located with the 40th Conference of the Spanish Society for Natural Language Processing (SEPLN 2024), CEUR-WS. org.

35. Wang, Y., Fang, Z., Du, W., Xu, S., Xu, R., & Li, J. (2024, May). Detection of Opioid Users from Reddit Posts via an Attention-based Bidirectional Recurrent Neural Network. In Proceedings of the 2024 8th International Conference on Medical and Health Informatics (pp. 318-324).

36. Yao, H., Rashidian, S., Dong, X., Duanmu, H., Rosenthal, R. N., & Wang, F. (2020). Detection of suicidality among opioid users on reddit: machine learning–based approach. Journal of medical internet research, 22(11), e15293.

37. Young, S. D., Lee, S. J., Perez, H., Gill, N., Gelberg, L., & Heinzerling, K. (2020). Social media as an emerging tool for reducing prescription opioid misuse risk factors. Heliyon, 6(3).

38. Laud, T., Kacha-Ochana, A., Sumner, S. A., Krishnasamy, V., Law, R., Schieber, L., ... & ElSherif, M. (2025, June). Large-scale analysis of online questions related to opioid use disorder on reddit. In Proceedings of the International AAAI Conference on Web and Social Media (Vol. 19, pp. 1068-1084).

39. Mittal, S., Jung, H., ElSherief, M., Mitra, T., & De Choudhury, M. (2025, June). Online myths on opioid use disorder: A comparison of reddit and large language model. In Proceedings of the International AAAI Conference on Web and Social Media (Vol. 19, pp. 1224-1245).

40. Rosenman, E., Buck, R. K., Anjum, N., Thompson, L., Rist, J., Banuna, L., ... & Holmes, L. M. (2025). From Dealer to Doctor: A Case Study Examining How Purdue Pharma Sought to Leverage Racial Health Disparities to Attenuate Flagging OxyContin Sales. Social Science & Medicine, 118132.

41. Frenzel, O., Ratz, I., Skarphol, A., & Werremeyer, A. (2025). Dare 2 Discuss Social Media Campaign: An educational initiative to improve Opioid Use Disorder discussions between the pharmacist and patient. Journal of the American Pharmacists Association, 102343.

42. Ahmad, M., Farid, H., Ameer, I., Ullah, F., Muzamil, M., Jalal, M., ... & Sidorov, G. (2025). UE-NER-2025: A GPT-based Approach to Multilingual Named Entity Recognition on Urdu and English. IEEE Access.

43. Ahmad, M., Batyrshin, I., & Sidorov, G. (2025). Sentiment Analysis Using a Large Language Model–Based Approach to Detect Opioids Mixed with Other Substances Via Social Media: Method Development and Validation. JMIR infodemiology, 5, e70525.

44. Yang, C., Chakrabarty, T., Hochstatter, K. R., Slavin, M. N., El-Bassel, N., & Muresan, S. (2023). Identifying Self-Disclosures of Use, Misuse and Addiction in Community-based Social Media Posts. arXiv preprint arXiv:2311.09066

45. Wang, Y., Fang, Z., Du, W., Xu, S., Xu, R., & Li, J. (2024, May). Detection of Opioid Users from Reddit Posts via an Attention-based Bidirectional Recurrent Neural Network. In Proceedings of the 2024 8th International Conference on Medical and Health Informatics (pp. 318-324).

46. Fodeh, S. J., Al-Garadi, M., Elsankary, O., Perrone, J., Becker, W., & Sarker, A. (2021). Utilizing a multi-class classification approach to detect therapeutic and recreational misuse of opioids on Twitter. Computers in biology and medicine, 129, 104132.

47. Lu, J., Sridhar, S., Pandey, R., Hasan, M. A., & Mohler, G. (2019). Redditors in recovery: text mining reddit to investigate transitions into drug addiction. arXiv preprint arXiv:1903.04081.

48. Kwon, S., Wang, X., Liu, W., Druhl, E., Sung, M. L., Reisman, J. I., ... & Yu, H. (2024, June). ODD: a benchmark dataset for the natural language processing based opioid related aberrant behavior detection. In Proceedings of the conference. Association for Computational Linguistics. North American Chapter. Meeting (Vol. 2024, p. 4338).

49. Laud, T., Kacha-Ochana, A., Sumner, S. A., Krishnasamy, V., Law, R., Schieber, L., ... & ElSherif, M. (2025, June). Large-scale analysis of online questions related to opioid use disorder on reddit. In Proceedings of the International AAAI Conference on Web and Social Media (Vol. 19, pp. 1068-1084).

50. Ahmad, M., Sidorov, G., Amjad, M., Ameer, I., & Batyrshin, I. (2025). Opioid Crisis Detection in Social Media Discourse Using Deep Learning Approach. Information, 16(7), 545.